\documentclass[graybox]{svmult}

\usepackage{mathptmx}       
\usepackage{helvet}         
\usepackage{courier}        
\usepackage{type1cm}        
\usepackage{makeidx}         
\usepackage{graphicx}        
\usepackage{multicol}        
\usepackage[bottom]{footmisc}

\usepackage[authoryear,round]{natbib}
\usepackage{graphicx}
\usepackage{soul}
\usepackage{hyperref}

\RequirePackage{amsmath,amsfonts,amssymb} 
\usepackage{booktabs,siunitx} 
\usepackage{tabularx} 
\usepackage{multirow}

\makeindex             

\begin{document}

\title*{Flutter stability of twin-box bridge decks}
\author{M. Rønne, A. Larsen, J. H. Walther and S. V. Larsen}
\institute{Maja Rønne \at COWI A/S, Parallelvej 2, 2800 Kgs. Lyngby, Denmark, \email{mjre@cowi.com} 
\and Allan Larsen \at COWI A/S, Parallelvej 2, 2800 Kgs. Lyngby, Denmark, \email{aln@cowi.com}
\and Jens H. Walther \at Technical University of Denmark, Nils Koppels Allé 403, 2800 Kgs. Lyngby, Denmark,\\ \email{jhwa@dtu.dk}
\and Søren V. Larsen \at FORCE Technology, Hjortekærsvej 99. 2800 Kgs. Lyngby, Denmark,\\ \email{svl@forcetechnology.com}}

%
\maketitle

\vspace{-3.5cm}
\abstract{The present paper reports on wind tunnel tests and analyses carried out to investigate the effect of the static angle of attack on the aerodynamic stability of a twin-box bridge deck section. It is found that the critical wind speed for onset of flutter increases with increasing positive static angles (nose-up) and that this effect relates mainly to a decrease in the loss of aerodynamic stiffness. 
A simplified flutter analysis, linking the slope of the static moment coefficient to the increase of flutter stability for increasing positive angles of attack.
It is concluded that it is desirable to design twin-box bridge deck sections to have a positive moment coefficient at zero angle of attack and a positive, decreasing moment slope for increasing nose-up angles. With these requirements fulfilled, the present study show that the critical wind speeds for onset of flutter increase with increasing angles, and ensures that the elastically supported deck will always meet the wind at ever increasing angles for increasing wind speeds.}


\section{Introduction}\label{sec:1}
The twin-box bridge deck, composed of two individual hull girders separated by a central air gap, have proven to have superior structural and aerodynamic performances as compared to a mono-box girder of same overall width of the carriage ways. 
For the structural perspective, the increase in lateral stiffness of the twin deck, with nearly the same linear mass as the equivalent mono-box girder, is a critical aspect for very-long span bridges. For the aerodynamic point of view, the increase in aeroelastic stability threshold, due to smaller slopes of the lift and moment coefficient, 
leads to increase in the instability wind speed \citep{Zasso2019}. 
The twin-box girder thus allows higher critical wind speeds for onset of flutter to be reached, as compared to the mono-box girder, and has therefore gained increasing popularity the recent years for design of super long span suspension bridges.

During the past decades experimental and numerical investigations on the aerodynamic performance of twin-box girders have widely been conducted, investigating the geometry of the twin girder and the width of the air gap (\cite{ Qui2007, Yang2015b, Yang2015, deMiranda2015, Nieto2020, Argentini_Rocchi}, just to mention some). 
\cite{Yang2015b,Yang2015, Qui2007} showed that an increase in central air gap (until a certain point) improved the flutter stability, as the torsional damping ratio increased, but that the aeroelastic effects also highly depended on the shape of the girder. \cite{Kwok2012} observed that the lift force and pitching moment were more sensitive to the angle of wind incidence than to the gap width. 
Further investigations of the angle dependency for the flutter derivatives for mono and multi box bridge girders are presented in \cite{Diana2004} and \cite{Mannini2008}. 

Recently, an analytical expression to determine the flutter derivatives for multi-box girders based on superposition of the flat plate theory has been introduced by \cite{Styrk_Eriksen}. Their results for the flutter derivatives show reasonable agreement with experimentally estimated flutter derivatives for $H_1^*$, $H_2^*$, $H_3^*$, and $A_2^*$. However, the torsional aerodynamic stiffness from the superposition of the flat plate is underestimated, which causes the model to overestimate the critical wind speed for flutter as compared to the free vibration test. \cite{Russo_ICWE16:2023} have modified the superposition of the flat plate, which results in fair agreement with experiments. Specially the aerodynamic torsional stiffness was improved being more accurate as compared to the superposition of the flat plate. 

During wind tunnel tests for the design of the twin-box girder for the 1915 Çanakkale Bridge, it was discovered that the critical wind speed for onset of flutter increased significantly when the elastically sprung deck section was allowed to rotate to positive angles of attack relative to the horizontal wind. A behaviour referred to as the “nose-up” effect. A theoretical investigation \citep{Ronne2021} supported the “nose-up” effect and linked it to the progression of the section moment and lift coefficient. 

The present paper reports on an experimental investigation carried out to support and quantify the “nose-up” effect for a twin-box bridge deck section similar to that of the 1915 Çanakkale Bridge. Furthermore, a simple flutter analysis is proposed, based on the Selberg formulation \citep{Larsen:2016}, and relates the critical flutter wind speed to the moment slope.

\section{Wind tunnel tests}
\begin{figure}
    \centering
    \includegraphics[width=.95\textwidth]{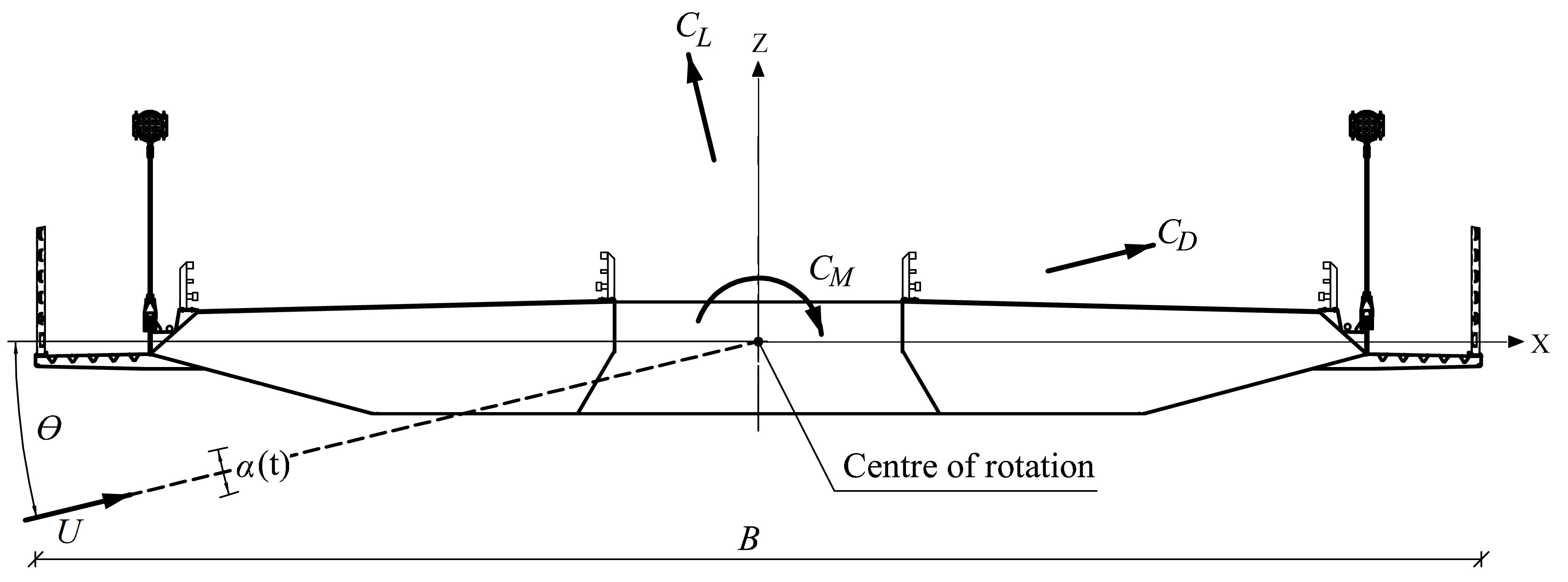}
    \caption{Geometry of the twin-box girder section tested at FORCE Technology and the definition of the sign convention for the static force coefficients. Note that for the flutter derivatives tests (forced motion tests) the z-axis is defined positive downwards, so in opposite direction as shown here, following Scanlan's original notation \citep{Scanlan_Tomko, Scanlan_Beliveau_Budlong}.}
    \label{fig:Force_secion_sign} 
\end{figure}

\begin{figure}
    \centering
    \includegraphics[width=.95\textwidth]{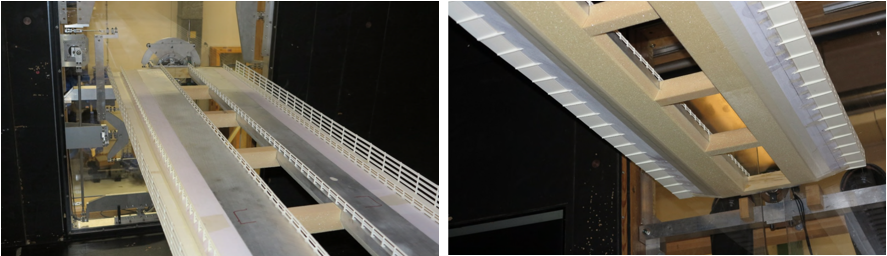}
    \caption{Top view of the section model in the forced motion rig in the wind tunnel (left), and bottom view of the model in the static rig (right).}\label{Fig:Force_WT}
\end{figure}

Wind tunnel tests were carried out for a twin-box section having hull girders of similar geometry as the 1915 Çanakkale Bridge, with cantilevered inspection walkways and 9 m wide central gap. Minor details such as the layout of wind screens and the railings are different. Also, the gantry rails running along the bottom plate of the 1915 Çanakkale Bridge are omitted in the present model. The cross section of the tested model is shown Fig. \ref{fig:Force_secion_sign} as well as the sign convention for the static force coefficient. Photos of the model setup in the forced motion rig and static rig in the wind tunnel are shown in Fig. \ref{Fig:Force_WT}. 

Static and forced motion tests were carried out at FORCE Technology's (2.6 m wide times 1.8 m high) Boundary Layer Wind Tunnel II, Kgs. Lyngby, Denmark. The tests were performed in smooth flow condition with along-wind and across-wind turbulence intensities ($I_u$ and $I_w$) of about $0.6\% - 0.7\%$. A 2.55 m long rigid section model of the twin-box girder in geometric scale $\lambda_L=1:80$ and with a chord $B = 0.569\, \mathrm{m}$ was used in the tests. 
The structural data of the 1915 Çanakkale Bridge reported in \autoref{T:Canakkale_data} is considered in the analyses of the test-case.  

%
\begin{table}[!htb]
\footnotesize
\caption{Structural data for the 1915 Çanakkale Bridge.}\label{T:Canakkale_data}
\centering
\begin{tabular}{ccccc}
\hline
Mass & Mass moment of inertia & Vertical freq. & Torsion freq. & Struc. damping \\
$m$ [$\text{kg/m}$] & $I$ [$\text{kg m}^2/\text{m}$] & $f_h$ [Hz] & $f_\alpha$ [Hz] & $\zeta$ [-]\\
\midrule
$28.85\cdot10^3$ & $6.215\cdot10^6$ & $0.072$ & $0.146$ & $0.003$\\
\hline
\end{tabular}
\end{table}

\subsection{Static force coefficient}
The static (time averaged) wind load coefficient for drag, lift and moment for the twin-box deck section were measured for wind angles of attack, $\theta$, between -10$^\circ$ to +10$^\circ$ at increments of $1^\circ$. 
For measuring the static force coefficients, the model in the static test rig were equipped with two 3-component force balances measuring the vertical, lateral and torsional reactions at the extremities of the model. 
The measurement of the static force coefficients was repeated for different wind speeds to verify their Reynolds number independence. The chosen wind speed, $U$, was between $8.5\, \mathrm{m/s}$ to $12.5\, \mathrm{m/s}$ depending of the angle of attack, corresponding to $\mathrm{Re}=\frac{\rho U B}{\mu} = 3-4.7\cdot 10^5$, where $\rho$ is the air density and $\mu$ the dynamic viscosity. 

The static force coefficients are defined as:
\begin{equation}\label{Eq:SFC}
C_D = \frac{D}{\frac{1}{2}\rho U^2 B}\,\, , \,\, C_L = \frac{L}{\frac{1}{2}\rho U^2 B}\,\, , \,\, C_M = \frac{M}{\frac{1}{2}\rho U^2 B^2}
\end{equation}
where $D$, $L$ and $M$ are the drag, lift and moment per unit length with the sign convention as shown in Fig. \ref{fig:Force_secion_sign} (moment positive nose-up, lift positive upwards). The  forces are normalized by the total deck width chord $B$, the air density $\rho$ and the mean wind velocity $U$.

The derivatives of the static force coefficients as function of the angle of attack $\theta$ are given as:
\begin{equation}\label{Eq:dSFC}
    K_{D,L,M} = \frac{d C_{D,L,M}}{d \theta}
\end{equation}

\begin{figure*}
    \centering
    \includegraphics[width=\textwidth]{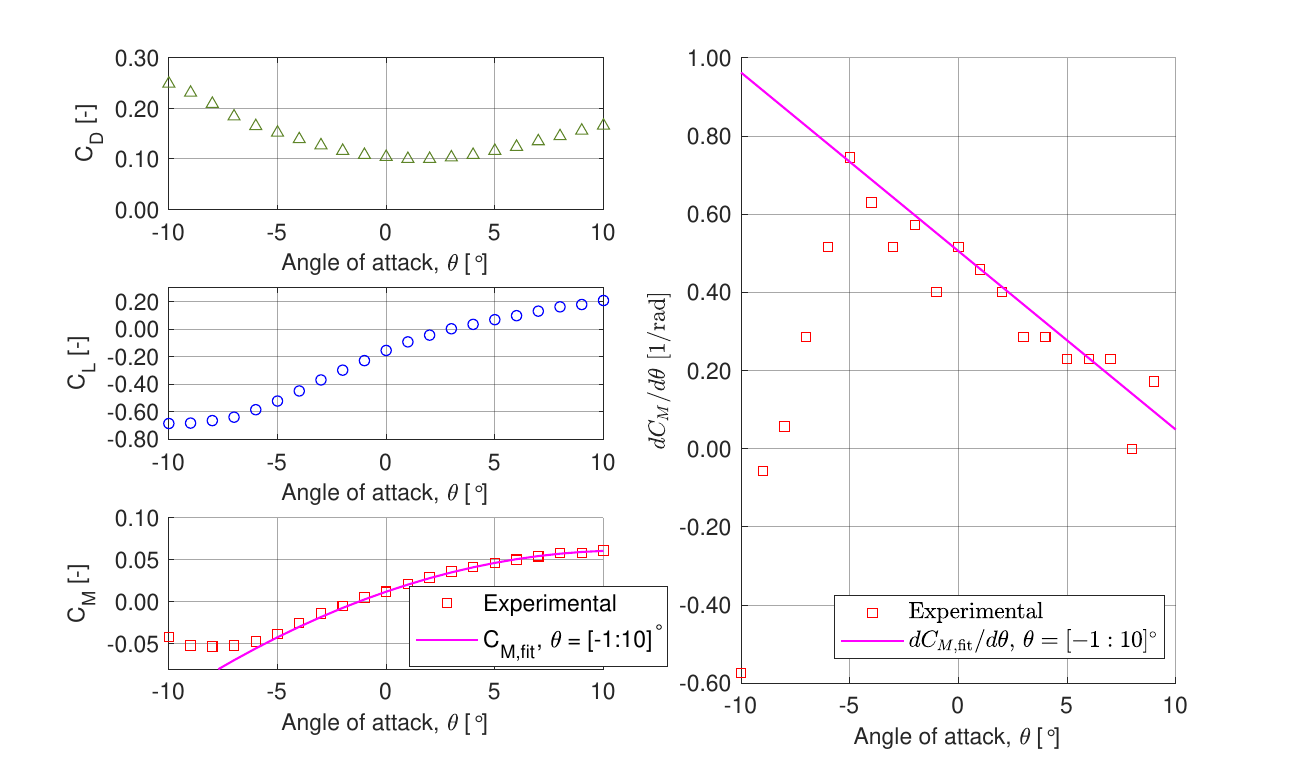}
    \caption{Static force coefficients $C_D$, $C_L$, $C_M$ and the moment slope $K_M=dC_M/d\theta$ measured from the wind tunnel tests. A second order polynomial fit to $C_M$ in the angular range of $\theta = [-1:10]^\circ$ is shown as well as the differentiated fit to $C_M$.}
    \label{fig:SFC_WTT}
\end{figure*}

The static force coefficients from the wind tunnel tests and the moment slope $K_M=\frac{d C_M}{d \theta}$ are shown in Fig. \ref{fig:SFC_WTT}.
It is noted that the moment coefficient at $\theta = 0^\circ$ angle of attack is positive, $C_{M0} = 0.012$, and that the moment slope is positive and continuously decreases for positive increasing angles. A second order polynomial is fitted to $C_M$ for the angle interval $\theta=[-1^\circ;10^\circ]$ to facilitate theoretical analysis (magenta curve in Fig.\ref{fig:SFC_WTT}).
The curve fit to $C_M$ is given in Eq. (\ref{Eq:CM-fit}) and the values are reported in \autoref{tab:CM-fit}. 

\begin{equation}\label{Eq:CM-fit}
    C_{M,\mathrm{fit}}(\theta)= C_{M2}\theta^2+C_{M1}\theta + C_{M0}
\end{equation}

\begin{table}[!htb]
    \footnotesize
    \caption{Parameter values for curve fit to $C_M$ from the experimental wind tunnel tests.}
    \label{tab:CM-fit}
    \centering
    \begin{tabular}{cccc}
    \toprule
        Fit & $C_{M2}$ & $C_{M1}$ & $C_{M0}$ \\
    \midrule
        $\theta$ in deg & $-3.985\cdot10^{-4}$ & $8.827\cdot10^{-3}$ & $0.012$\\
        $\theta$ in rad & $-1.308$ & $0.5057$ & $0.012$\\
    \bottomrule
    \end{tabular}
\end{table}

\subsubsection{Section rotation to mean wind speed}
The development of the angle of attack $\theta$ as function of mean wind speed $U$ can be evaluated by assuming that the aerodynamic moment is balanced by the moment capacity of the elastically suspended section. 
The simplified static moment equilibrium for a 1DOF (degree-of-freedom) oscillator model is then: 
\begin{equation}\label{Eq:Iomega_a}
    I\omega_\alpha^2\theta=\frac{1}{2}\rho U^2B^2 \left(C_{M2}\theta^2+C_{M1}\theta + C_{M0} \right)
\end{equation}
where $I\omega_\alpha^2$ is the torsional static stiffness of the system with $I$ beeing the mass moment of inertia per unit length, $\omega_\alpha$ is the natural oscillation frequency of the torsion motion, $C_{M2}$, $C_{M1}$ and $C_{M0}$ are the parameter values for the curcve fit to the moment coefficient.
%
The solution to Eq. (\ref{Eq:Iomega_a}) is obtained as:

\begin{equation}\label{Eq:alpha_U_2order}
    \theta(U) = \frac{-\left(C_{M1}-\frac{I\omega_\alpha^2}{\frac{1}{2}\rho U^2B^2}\right)\pm \sqrt{\left(C_{M1}-\frac{I\omega_\alpha^2}{\frac{1}{2}\rho U^2B^2}\right)^2-4C_{M2}C_{M0}}}{2C_{M2}}
\end{equation}

\begin{figure}
    \centering
    \includegraphics[width=.65\linewidth]{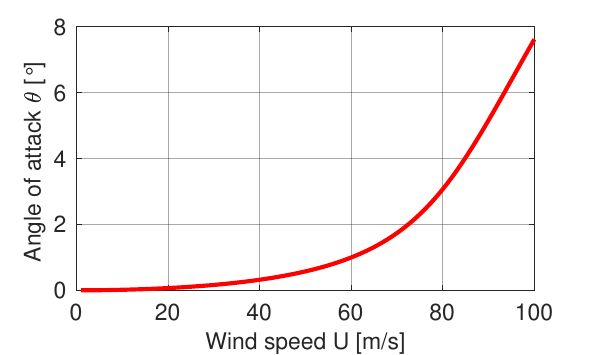}
    \caption{Development of static angle of attack $\theta$ as function of mean wind speed $U$.}
    \label{fig:Angle_Ur}
\end{figure}

Taking the first symmetric torsion mode of the 1915 Çanakkale Bridge from \autoref{T:Canakkale_data} and using the moment fit parameter values from \autoref{tab:CM-fit} in Eq. (\ref{Eq:alpha_U_2order}), yields a development of the angle of attack $\theta$ shown in Fig. \ref{fig:Angle_Ur}. The development of the angle $\theta$ as function of wind speed $U$ is dependent on the development of the $C_M$. 
It is noted that the development of the $C_M$ curve (see Fig. \ref{fig:SFC_WTT}) makes the angle of attack increase as function of mean wind speed. At first slowly but at steeper rates for wind speeds above $60\, \mathrm{m/s}$. For a mean wind speed $U = 45\, \mathrm{m/s}$ which is typical for design of large suspension bridge in Europe, the angle of attack of the deck will be about $0.5^\circ$ which is about 1/3 of the typical cross fall of the roadway. 
At a mean wind speed of $90\, \mathrm{m/s}$ an angle of attack of about $5^\circ$ is estimated. 
It is important to note that the slope of the $C_M$ curve in Fig. \ref{fig:SFC_WTT} is steeper at $\theta=0^\circ$ than at $\theta=5^\circ$. This feature is influencing the development of the angle of attack as function of the wind speed.
A positive moment coefficient $C_M$ at $\theta=0^\circ$ angle of attack, $C_{M0}>0$, and a positive moment slope $\frac{d C_M}{d\theta}>0$, decreasing with increasing rotation angles, $\frac{d^2 C_M}{d \theta^2}<0 $ for $\theta>0$, ensures that the elastically supported deck will always meet the wind at ever increasing angles for increasing wind speeds.

\subsection{Aerodynamic flutter derivatives}
Vertical and torsion flutter derivatives were measured by the forced motion techniques (see \cite{SVLarsen_Sinding_Smitt}) for angles of wind incidence of $\theta = -2^\circ$, $0^\circ$, $+2^\circ$, and $+4^\circ$ relative to horizontal, for reduced wind speed up to $U_r = \frac{U}{fB}=25$. The model scale amplitudes were $\pm 8\, \mathrm{mm}$ for the vertical motion and $\pm 1^\circ$ for the rotation.

The self-excited aeroelastic forces acting on the bridge deck for a 2DOF flutter case, are expressed through the aerodynamic derivatives following Scanlan's originally definition \citep{Scanlan_Tomko}: 

\begin{equation}\label{Eq:L_ea}
    L_{ae}=\rho U^2 B \left[K H_1^* \frac{\Dot{h}}{U} + K H_2^* \frac{B\Dot{\alpha}}{U} + K^2 H_3^*\alpha + K^2 H_4^*\frac{h}{B}\right]
\end{equation}

\begin{equation}\label{Eq:M_ea}
       M_{ae}=\rho U^2 B^2 \left[K A_1^* \frac{\Dot{h}}{U} + K A_2^* \frac{B\Dot{\alpha}}{U} + K^2 A_3^*\alpha + K^2 A_4^*\frac{h}{B}\right]
\end{equation}
where $L_{ae}$ is aeroelastic lift force per unit span length (defined positive downwards) and $M_{ae}$ is the aeroelastic moment per unit span length (positive nose-up). Furthermore, $K=\frac{\omega B}{U}$ is the reduced frequency where $\omega$ is the circular oscillation frequency, $h$, $\alpha$, $\dot{h}$ and $\dot{\alpha}$ are the vertical heave and rotational pitch displacement and their corresponding time derivatives. $H_i^*$ and $A_i^*$ ($i=1,2,3,4$) are the aerodynamic flutter derivatives. 

\begin{figure*}
    \centering
    \includegraphics[width=.99\linewidth]{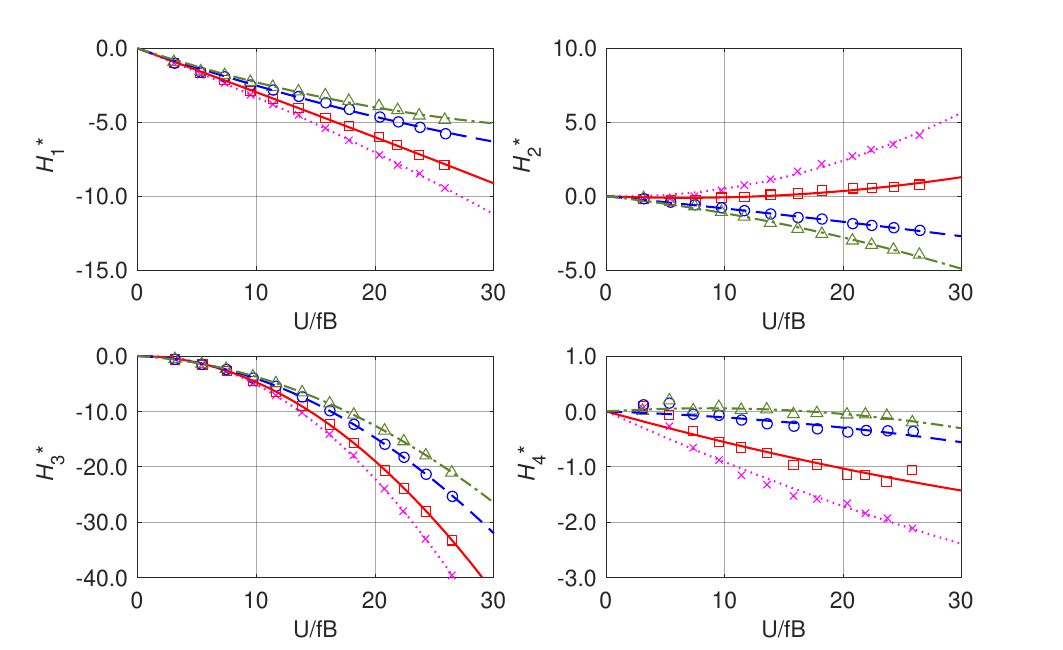}
        \includegraphics[width=.99\linewidth]{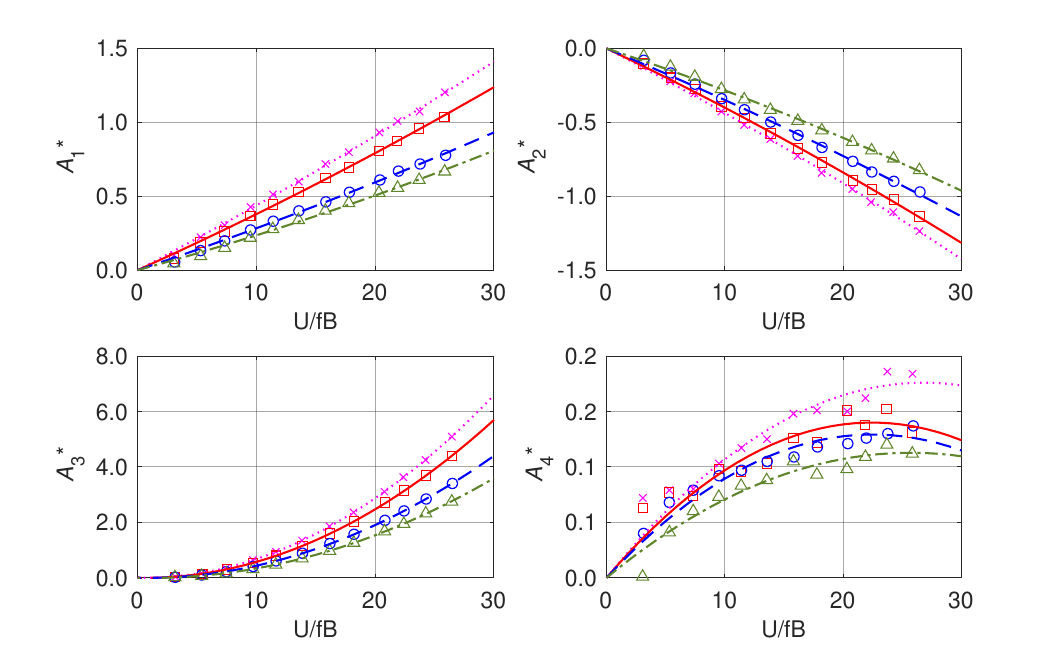}
        \includegraphics[width=.79\linewidth]{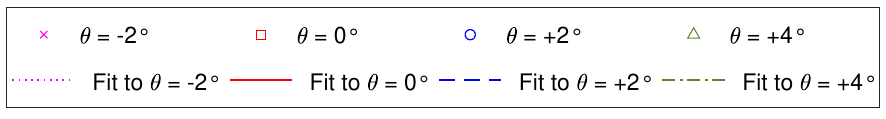}
    \caption{Aerodynamic flutter derivatives obtained in the wind tunnel as function of the reduced velocity for different angles of attack and curve fits of second order polynomial.}\label{fig:FD_WT}
\end{figure*}

The measured aerodynamic flutter derivatives as function of the reduced velocity for the different angles of attack are reported in Fig. \ref{fig:FD_WT}. Second order polynomials are fitted to the measured data for each angle. As seen from Fig. \ref{fig:FD_WT} the second order polynomials fits are in fair agreement with the measurements. 
The sign convention and normalization adopted for the aerodynamic derivatives is that originally proposed by \cite{Simmiu_Scanlan1985}.

From Fig. \ref{fig:FD_WT} it is indicated that the aerodynamic derivatives change in a systematic way as function of angle of attack. For a given reduced velocity, the magnitude of the aerodynamic derivatives seems to increase or decrease, depending of the angle of attack, and aerodynamic derivative of interest. The angle dependency for the flutter derivatives are also noted by \cite{Argentini_Rocchi} and seems to be related to the geometrical details for the twin-box girder as well.
In Fig. \ref{fig:FD_WT} it is seen that the $A_3^*$ aerodynamic derivative, related to the loss of aerodynamic stiffness, is decreasing for increasing angles. This observation is one of the main reasons that the bridge deck reaches higher critical wind speeds for onset of flutter for increasing positive angles of attack (see Fig. \ref{fig:Ucr_flutter_WT}).  

%

\section{Flutter analysis}
\subsection{The AMC method}
The aerodynamic derivatives obtained from the wind tunnel tests are applied in a flutter analysis for the twin-box deck section using the structural data listed in \autoref{T:Canakkale_data}. The critical wind speed for onset of flutter is determined by the Air Material Command (AMC) method \citep{Scanlan:1962, Smilg:1942} 
by balancing the apparent aerodynamic damping $g$ to twice the structural damping $2\zeta_s$ of the bridge. The AMC method requires that the structural damping is identical for all considered modes, and it is assumed that the structural damping forces are proportional to the displacement but phase shifted $90^\circ$. In this way the flutter problem is transformed into a complex eigenvalue problem.

\begin{figure}
    \centering
    \includegraphics[width=.75\linewidth]{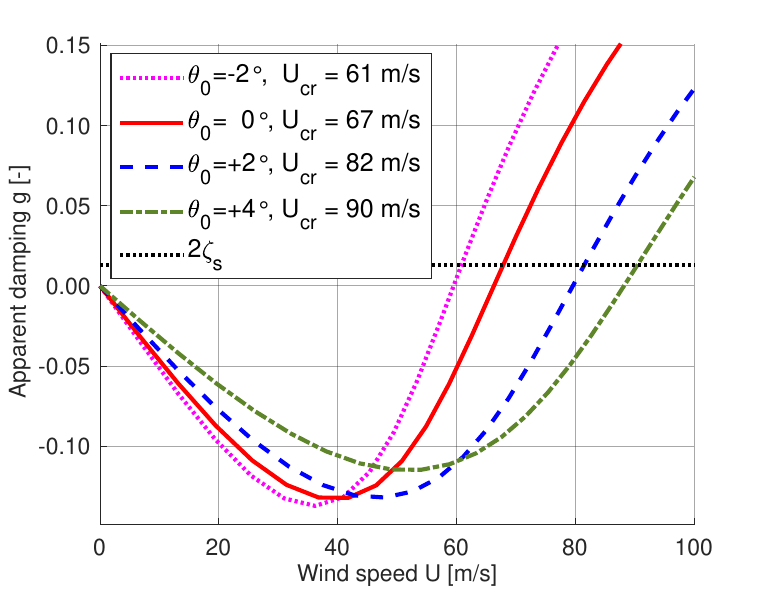}
    \caption{Flutter analysis for the twin-box deck section by the AMC method. The critical wind speed for onset of flutter is determined as the intersection of the apparent aerodynamic damping curves $g$ and twice the structural damping $2\zeta_s$.}
    \label{fig:Ucr_flutter_WT}
\end{figure}

The AMC method is therefore well suited for analysis involving multiple modes. The method is outlined in details in \cite{Ronne2021}. 
The results of the flutter analysis by the AMC method, yielding predicted critical wind speeds for onset of flutter, are shown in Fig. \ref{fig:Ucr_flutter_WT} for the different angles of attack. The critical wind speeds are summarized in \autoref{T:Ucr}. It is observed that the critical wind speed for onset of flutter increases for increasing angles of attack.

\subsection{Simplified flutter analysis}
Simple formulas for estimation of the critical wind speed for onset of flutter are convenient as for initial estimates during early stages of a long span bridge project. One such equation borrowing from quasi-steady theory relates the slope of the moment coefficient $d C_M/d \theta$ to the critical wind speed for onset of flutter $U_{cr}$:

\begin{equation} \label{eq:Ucr} 
    U_{cr}= 2\pi f_\alpha B\sqrt{\frac{I \left(1-\left(\frac{f_h}{f_\alpha}\right)^2 \right)}{\rho B^4 \frac{\,d C_M}{\,d \theta} \mathcal{F}}}
\end{equation}

The simplified flutter equation for estimating the critical wind speed for onset of flutter, Eq. (\ref{eq:Ucr}), is a kin of Selberg's formula \citep{Larsen:2016}, but allows the variation of the critical wind speed as function of $\theta$ to be accounted for. The equation includes the empirical factor $\mathcal{F}$ applied to the moment slope that serves the same purpose as the real part of Theodorsen circulation function in the "flat plate" aerodynamics. The empirical factor $\mathcal{F}$ serves to model the depreciation of the aerodynamic moment caused by the oscillatory wake of the deck section. 

The simplified flutter estimation can be derived from the equation of motion for torsion, in a simplified form, by neglecting structural damping and assuming that the aerodynamic loading corresponds to the static aerodynamic moment.
It is assumed that $\omega^2=\frac{1}{2}(\omega_\alpha^2+\omega_h^2)$, which based on experience, is a fair estimate for the flutter frequency. The derivation of the simplified flutter equation is described in more detail \cite{LarsenIABSE:2022}. 
%

The expression for the critical wind speed for onset of flutter in Eq. (\ref{eq:Ucr}), it is noted that $U_{cr}$ increases for (positive) decreasing moment slope $\frac{dC_M}{d\theta}$. This observation supports the nose-up effect, that the critical wind speeds of flutter increase with increasing angle of attack, as the moment slope decrease with increasing angles.

Taking $\mathcal{F} = 0.73$ will make $U_{cr}$ calculated by Eq. (\ref{eq:Ucr}) 
coincide with the critical wind speed obtained from the flutter analysis based on the aerodynamic derivatives for $\theta = 0^\circ$ from the wind tunnel data. 
The value of $\mathcal{F} = 0.73$ is estimated based on an assmued critical flutter wind speed in the order of $60-80\, \mathrm{m/s}$ giving a critical reduced frequency in the order of $0.4-0.5$, see  \cite{Larsen:2016}.
Introducing the second order polynomial fit to $C_M$ from the wind tunnel tests ( Eq. (\ref{Eq:CM-fit}) and \autoref{tab:CM-fit}) in Eq. (\ref{eq:Ucr}), allows $U_{cr}$ to be estimated for other angles of attack. The result is listed in \autoref{T:Ucr} and compared to the flutter analysis derived from the AMC method, applying the measured aerodynamic derivatives. 

\begin{table}[!htb]
\footnotesize
\caption{Critical wind speeds for onset of flutter based on aerodynamic derivatives (AMC method) and simplified theory for the wind tunnel tests.}\label{T:Ucr}
\centering
\begin{tabular}{ccc}
\toprule
Angle of attack $\theta$ [$^\circ$] & AMC method $U_{cr}$ [m/s] & Simple formulae $U_{cr}$ [m/s] \\ 
\midrule
$-2$                & $61$              & $62$ \\
$0$                 & $67$              & $67$ \\
$+2$                & $82$              & $74$ \\
$+4$                & $90$              & $83$ \\
\bottomrule
\end{tabular}
\end{table}


From \autoref{T:Ucr} it is observed that the simplified flutter analysis, based on the static force coefficient measured from the wind tunnel tests, gives similar results for the critical flutter wind speed as the flutter analysis based on the measured aerodynamic derivatives. The simplified flutter analysis is in the present case slightly conservative for the positive angles of attack. For the case of $\theta=-2^\circ$ the flutter analysis based on the aerodynamic derivatives is estimating a critical wind speed of $U_{cr} = 61$ m/s as compared to $U_{cr} = 62$ m/s for the simplified flutter analysis based on the static moment coefficient slope, $\frac{dC_M}{d\theta} = 2C_{M2}\theta + C_{M1}$, from the wind tunnel tests.


\section{Conclusion}
The analysis of wind tunnel tests of a twin-box bridge deck section confirm that the flutter stability is dependent on the cross sections static angle to the wind, yielding higher critical wind speeds for higher positive (nose-up) angles of attack. A simplified expression linking the critical wind speed for onset of flutter to the slope of the moment coefficient yields a similar trend although the effect is less pronounced than the critical wind speed obtained by a full flutter analysis including 8 aerodynamic derivatives. The present experimental study thus confirms earlier theoretical analyses and findings. 

Considering the geometrical layout of twin-box bridge decks for maximum aerodynamic stability, the present study confirms that it is desirable to design for positive (nose-up) aerodynamic moment at zero angle of attack and a positive, decreasing moment slope for increasing angles.

Based on the flutter analysis of the aerodynamic derivatives, it is found that the increase of the critical wind speed for onset of flutter with increasing angles of attack, mainly relates to the decrease in the loss of aerodynamic stiffness with increasing in flow angles of attack.
\bibliographystyle{cas-model2-names}

\bibliography{cas-refs}
\end{document}